\documentclass[aps, twocolumn, showpacs]{revtex4}
\usepackage{graphicx}
\usepackage{amsmath,amssymb}
\pdfoutput=1

\begin{document}

\title{Dynamics of particle flips in two-dimensional quasicrystals}

\author{Michael Engel}
\email{engelmm@umich.edu}
\affiliation{Department of Chemical Engineering, University of Michigan, Ann
  Arbor, MI 48109-2136, USA}
\author{Masahiro Umezaki}
\affiliation{Department of Physics, Kyushu University, Fukuoka 812-8581,
  Japan\footnote{Current address: Japan Meteorological Agency, Tokyo 100-8122,
    Japan}}
\author{Hans-Rainer Trebin}
\affiliation{Institut f\"ur Theoretische und Angewandte Physik, Universit\"at
  Stuttgart, Pfaffenwaldring 57, 70550 Stuttgart, Germany}
\author{Takashi Odagaki}
\affiliation{School of Science and Engineering, Tokyo Denki University,
  Hikigun Hatoyama, Saitama 350-0394, Japan}

\date{\today}

\begin{abstract}
The dynamics of quasicrystals is more complicated than the dynamics of
periodic solids and difficult to study in experiments. Here, we
investigate a decagonal and a dodecagonal quasicrystal using molecular
dynamics simulations of the Lennard-Jones-Gauss interaction system. We
observe that the short time dynamics is dominated by stochastic
particle motion, so-called phason flips, which can be either
single-particle jumps or correlated ring-like multi-particle
moves. Over long times, the flip mechanism is efficient in reordering
the quasicrystals and can generate diffusion. The temperature
dependence of diffusion is described by an Arrhenius law. We also
study the spatial distribution and correlation of mobile particles by
analyzing the dynamic propensity.
\end{abstract}


\maketitle

\section{Introduction}

\label{Introduction}
Quasicrystals are long-range ordered structures without
periodicity. They are traditionally found in metallic
alloys~\cite{DiscoveryQC} and recently also in other
materials~\cite{PolymerQC,ColloidQC,TetrahedraQC}. The dynamics of
quasicrystals is characterized by two elementary excitations, phonons
and phasons~\cite{Primer}. As in periodic solids, phonon modes
correspond to the oscillatory motion of atoms around their equilibrium
positions. In contrast, phasons describe local rearrangements of
atoms, which are present due to the absence of periodic
order. Possible rearrangements are restricted by the internal geometry
of the quasicrystal. On the continuum level, phason modes are a
concept of the hydrodynamic theory of
quasicrystals~\cite{BakHydro,LubenskyHydro,QC_basis}; in contrast to
phonon modes, they are diffusive in nature and can only be excited
internally via phason-phonon interactions. Long wave length phason
fluctuations are present in thermodynamic equilibrium as evidenced by
diffuse neutron scattering~\cite{PhasonFluctuations}.

On the atomistic scale, the elementary process of a phason mode is a
phason flip, which is the individual motion of a single atom or the
correlated multi-atom motion of a few atoms. A phason flip transforms
one local configuration into a similar, energetically nearly
degenerate one by overcoming an energy barrier. The concept of phason
flips plays an important role in explaining enhanced atomic
diffusion~\cite{KKmodel,TracerStudy}, dislocation
motion~\cite{dislocations}, and structural phase
transitions~\cite{structuralphase1,structuralphase2} in
quasicrystals. To analyze the flips on a microscopic level, various
experimental works have been carried out. With the use of transmission
electron microscopy images collective phason flips were observed {\it
  in-situ} as rearrangements of atomic
clusters~\cite{edagawa-TEM}. Indirect evidence for phason flips was
obtained from the observation of phason walls traced out by
dislocations~\cite{dislocationObservation}. Furthermore, quasi-elastic
neutron scattering is believed to contain information about phason
flips~\cite{CoddensEXP}, although the contribution of the flips is
hard to extract because it is hidden in the elastic peak. Despite the
effort, phason modes and phason flips remain difficult to study in
experiment and little is known about the underlying processes on an
atomistic level.

Numerical studies of quasicrystal formation and dynamics were
performed using model potentials such as the binary Lennard-Jones (LJ)
potential in two dimensions~\cite{FirstMD2} and three
dimensions~\cite{IcoQC3D}, and the one-component Dzugutov
potential~\cite{Dzugutov} among other
systems~\cite{StrucComplexity}. In the binary LJ system, the ratio of
the particle radii are chosen to favor decagonal (2D) and icosahedral
(3D) order. Similarly, the potential of the Dzugutov system is
characterized by two competing length scales. Distances around the
potential maximum are strongly disfavored. By positioning the maximum
at the characteristic interparticle distances of the triangular
lattice or fcc/hcp, these simple close-packed lattices are
destabilized and the particles are forced to find other, more complex
configurations. Although quasicrystals have now been observed in
several model systems as thermodynamically stable
phases~\cite{StrucComplexity}, the energetic ground state of all
systems studied in simulation so far is crystalline. Examples are
two-dimensional monodisperse systems with a square
potential~\cite{SquarePot} or the binary LJ
system~\cite{QCgroundState}. A quasicrystalline structure exists in
these systems only at finite temperature and is therefore stabilized
entropically.

Oscillatory interaction potentials are often used for quasicrystal
simulations, because effective pair potentials for metals, which are
by far the most important materials class for quasicrystals, are best
described by a strongly repulsive part and a decaying oscillatory
(Friedel,~\cite{Friedel}) term.  Classical pair potentials fitted to
reproduce ab-initio forces typically show an oscillatory
behavior~\cite{MoriartyWidom,MD_qc}. To investigate the influence of
competing energy minima in the potential, Engel and Trebin introduced
the Lennard-Jones-Gauss (LJG) system~\cite{LJG}, which consists of
identical particles interacting with a parameterized double-well pair
potential. Its phase diagram as a function of the potential shape is
complicated and includes at least two entropically stabilized
quasicrystalline phases together with around a dozen of crystalline
periodic phases~\cite{StrucComplexity,Engel_Dpaper}. The LJG system
therefore constitutes a simple model system for studying quasicrystals
in simulation.

The aim of the present paper is to investigate phason flips using
molecular dynamics simulations. In particular, we are interested in
the basic mechanism on the particle level and the temperature
dependence of the dynamics. Our model system consists of identical
particles interacting in two dimensions with the LJG pair
potential. The paper is organized as follows: Sec.~\ref{Model}
introduces the LJG system and numerical simulation methods. In
Sec.~\ref{sec:form_qc}, the formation and thermodynamics of a
decagonal and a dodecagonal quasicrystals are examined.
Sec.~\ref{statistics} investigates individual and collective particle
flips and discusses their role for diffusion. In
Sec.~\ref{propensity}, we study the spatial distribution of mobile
particles from the dynamic propensity. Sec.~\ref{conclusion} concludes
with a brief summary.

\section{Model system and methods}\label{Model}

\subsection{Lennard-Jones-Gauss potential}

The system under investigation consists of identical particles interacting
with the LJG potential~\cite{LJG}
\begin{align}\label{eq:LJG}
V(r) = \varepsilon _0 \left \{ 
\left ( \dfrac{r_0}{r}\right )^{12} 
-2 \left ( \dfrac{r_0}{r}\right )^{6} 
-\varepsilon \exp \left ( -\dfrac{(r-r_G)^2}{2r_0^2\sigma^2} \right ) 
\right \},
\end{align} 
where $\varepsilon _0$ is the energy unit and $r_0$ the length unit.
$r_G$, $\varepsilon$, and $\sigma^2$ are three potential
parameters. The first two terms in Eq.~(\ref{eq:LJG}) are the
well-known LJ potential that remains fixed. The third term is a
Gaussian well ($\varepsilon>0$), whose center, depth, and width are
$r_G$, $\varepsilon$, and $\sigma$, respectively. Depending on the
choice of parameters, the LJG potential can be either a double-well
(Fig.~\ref{fig:pote_dec}, top left) or single well with shoulder for
small $r_0$ (Fig.~\ref{fig:pote_dodec2}, top left).

\subsection{Molecular dynamics simulation}

Molecular dynamics simulations are carried out with a system of 1024
particles using periodic boundary conditions. We solve the equations
of motion with the leapfrog algorithm in the NPT ensemble employing a
Nos\'e-Hoover thermostat for temperature control and an Andersen
barostat for pressure control. Throughout the paper the pressure is
fixed at $P=0$, which means we relax the boundaries such that the
potential energy remains minimal. The particles are arranged randomly
in the initial configuration with their velocities chosen according to
a Maxwell-Boltzmann distribution. Simulation units are dimensionless:
the length unit is $r_0$, the temperature unit is $\varepsilon _0
/k_B$, and the time unit is $\tau=\sqrt{mr_0^2/\varepsilon _0}$. Here,
$k_B$ is the Boltzmann constant and $m$ the particle mass. A single
molecular dynamics time step is equal to $0.01\tau$. The potential
cutoff is set to $r_{\mathrm{cut}}/r_0=2.5$. Beyond $r_{\mathrm{cut}}$
the LJG potential is essentially zero for the potential parameters
under investigation.

\subsection{Correlation functions}

The mobility of the $j$-th particle with trajectory $\mathbf{r}_j(t)$
is given by the dynamic propensity~\cite{propensity1,propensity2}
\begin{align}\label{eq:propensity} 
\phi_j(t) = \langle \Delta \mathbf{r}_j(t)^2 \rangle_{\text{ic}}
= \langle [\mathbf{r}_j(t)-\mathbf{r}_j(0)]^2 \rangle_{\text{ic}}.
\end{align}
The angle brackets indicate the average over an iso-configurational
ensemble, i.e.\ the simulation repeatedly starts from the same
particle configuration but with momenta randomly assigned from the
Maxwell-Boltzmann distribution. By averaging over the particles, we
obtain the mean square displacement
\begin{align}
\langle r^2(t)\rangle =
\frac{1}{N}\sum_{j=1}^N\langle [\mathbf{r}_j(t)-\mathbf{r}_j(0)]^2 \rangle,
\end{align}
which is related to the diffusivity $D$ of a $d$-dimensional system by the
Einstein equation
\begin{align}
D=\frac{1}{2d}\lim_{t\rightarrow\infty}\frac{\langle r^2(t)\rangle}{t}.
\end{align}

The radial distribution function $g(r)=\rho(r)/\rho$ is the average
density at distance $r$ divided by the global average density, where
\begin{align}
\rho(\mathbf{r})=\dfrac{1}{N}\sum_{j=1}^N \langle \delta(\mathbf{r} - 
\mathbf{r}_j(0))\rangle, 
\end{align}
is the density distribution. The distance and direction of the motion
of an individual particle are measured by the van Hove autocorrelation
function~\footnote{The van Hove autocorrelation function is also
  called the `self' part or the incoherent part of the van Hove
  correlation function.}
\begin{align}\label{eq:auto_col}
G_{a}(\mathbf{r},t)=\dfrac{1}{N}\sum_{j=1}^N \langle \delta(\mathbf{r} - 
\mathbf{r}_j(t) + \mathbf{r}_j(0) )\rangle,
\end{align}
which has a peak at $\mathbf{r}=0$ and decays outwards. The static
structure factor
\begin{align}
S(\mathbf{q})=\dfrac{1}{N}\left\langle \left|\sum_{j=1}^N
\exp(-i\mathbf{q}\cdot\mathbf{r}_j(t))\right|^2\right\rangle_t
\end{align}
is the usual non-energy resolved diffraction image
$S(\mathbf{q})=|\rho(\mathbf{q})|^2$ as measured in diffraction
experiments to determine crystal structure and symmetry.

\subsection{Local structure analysis}

In the theory of quasicrystals the crystal structure is described as a
(decorated) tiling~\cite{Primer}. Each tile represents a certain local
configuration. Due to the aperiodic order of quasicrystals, there is
only a finite number of different tiles up to
translation. Two-dimensional tilings are especially easy to visualize,
because the tiles are polygons. In the quasicrystals found in the LJG
system, a simple tiling is defined by the network of nearest
neighbors: two particles are linked (i.e.\ nearest neighbors), if
their distance is within the first peak of the radial distribution
function. Individual tiles are identified by finding closed paths in
the neighbor network. Examples for tile types that will be relevant in
the following are triangles, squares, and pentagons. Polygonal tiles
can be either empty (three to six vertices) or contain a single
particle like for example the decagon discussed below. In general,
tiles can also be concave or have irregular shape.

\section{A decagonal and a dodecagonal quasicrystal}\label{sec:form_qc}

\begin{figure}
\begin{center}
\includegraphics[angle=-90,width=\columnwidth]{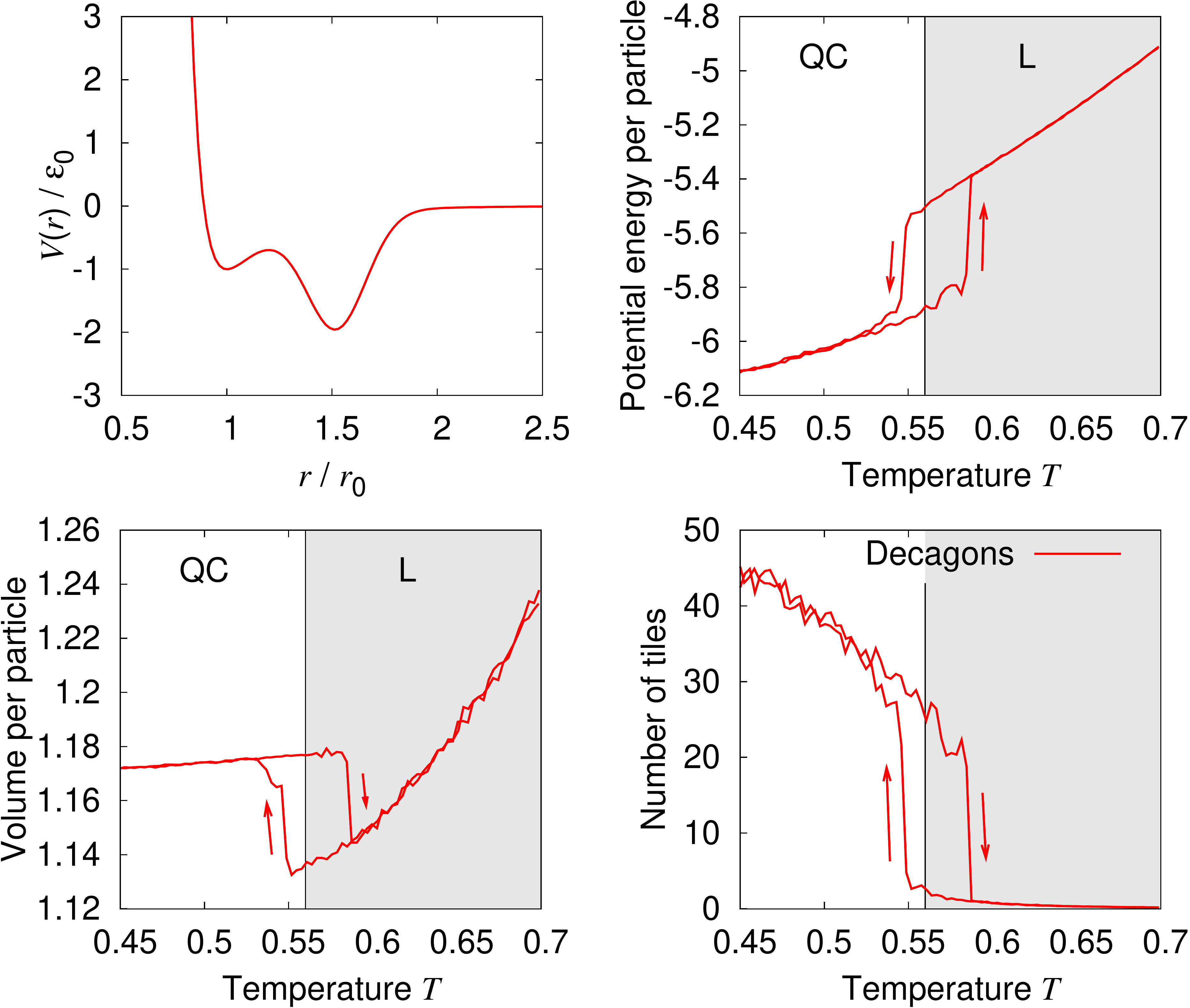}
\end{center}
\vspace*{-5mm}
\caption{(Top left)~The Lennard-Jones-Gauss potential for $r_G=1.52$,
  $\varepsilon =1.8$, and $\sigma^2=0.02$ is a double-well. (Top right)~The
  system exhibits a first order phase transition from quasicrystal~(QC) to
  liquid~(L) with (Bottom left)~negative thermal expansion. (Bottom right)~The
  number of decagons sharply increases at crystallization.}
\label{fig:pote_dec}
\end{figure}

We study two sets of potential parameters that stabilize a decagonal and a
dodecagonal quasicrystal in thermodynamic equilibrium. With the parameters
\begin{align}\label{parameter:dec}
r_G=1.52\text{, } \varepsilon = 1.8\text{, and } \sigma^2=0.02, 
\end{align} 
a phase transition from a decagonal quasicrystal to a liquid is found at
$T_{M1}=0.56\pm0.02$ (Fig.~\ref{fig:pote_dec}).  The transition is unusual,
because (i)~it is clearly first order and therefore does not resemble the
melting transition of the hexagonal crystal in the 2D LJ system, which is
characterized by the existence of an intermediate hexatic phase with unpinned
disclination pairs~\cite{2Dmelting}. No analog of a hexatic phase and no
unpinned disclination are found in the decagonal quasicrystal before
melting. (ii)~The transition is accompanied by negative thermal expansion
(Fig.~\ref{fig:pote_dec}, bottom left). During crystallization, the system
expands by 3\%. A similar behavior is known for example in water. As expected,
the compressibility of the crystal is much lower than the compressibility of
the liquid.

\begin{figure}
\begin{center}
\includegraphics[width=0.48\columnwidth]{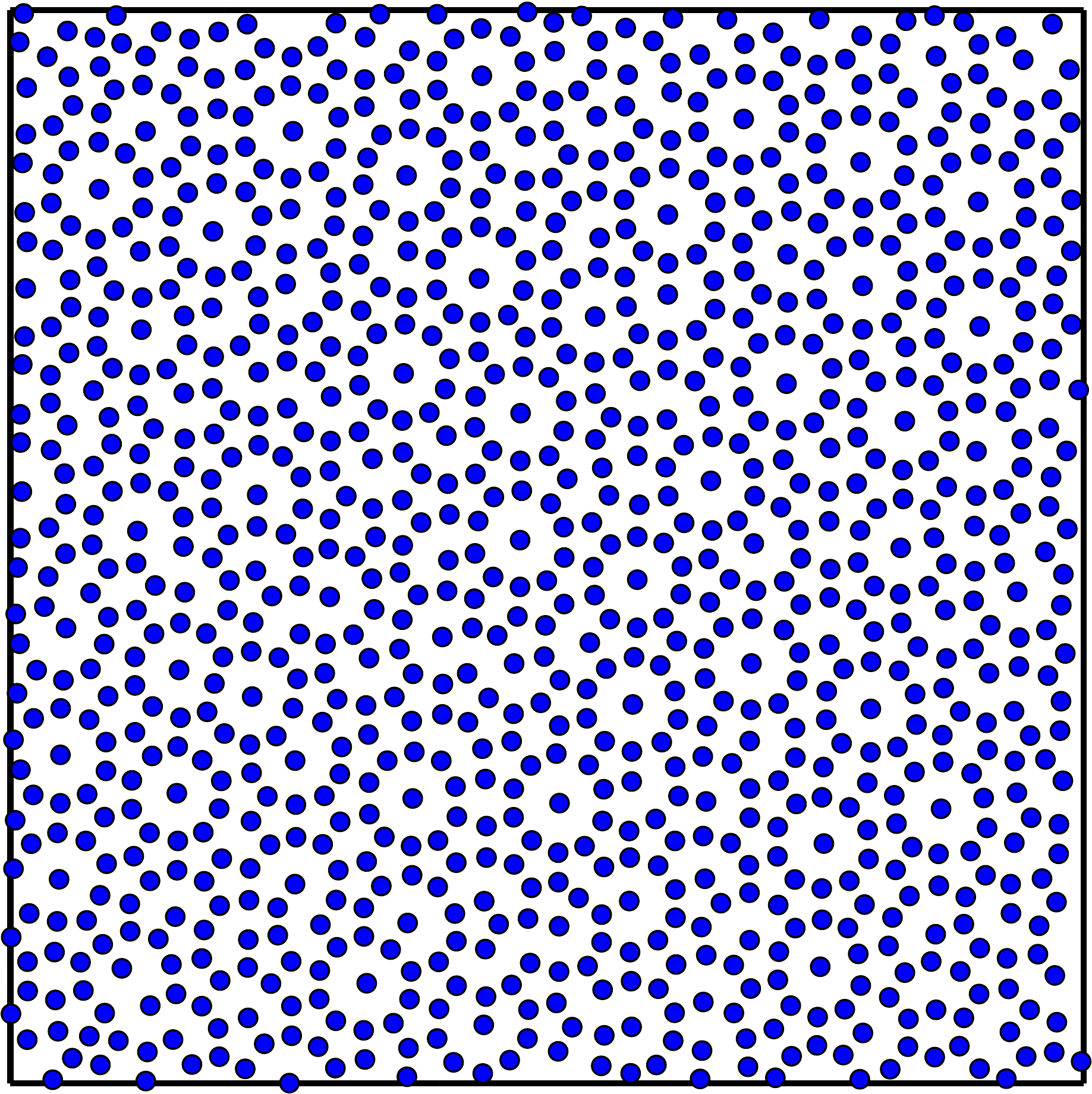}\quad
\includegraphics[width=0.48\columnwidth]{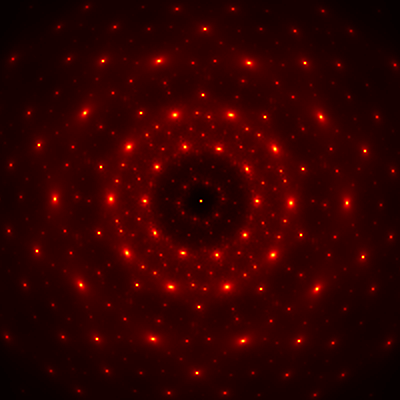}
\end{center}
\vspace*{-5mm}
\caption{Decagonal phase at $T=0.9T_{M1}=0.5$. (Left)~Particle configuration
  in real space. (Right)~Intensity map of the static structure factor.}
\label{fig:dec_phase_img}
\end{figure}

The reason for the negative expansion is the open structure of the decagonal
quasicrystal (Fig.~\ref{fig:dec_phase_img}, left). Especially decagon tiles (a
central particle surrounded by a ring of ten particles) are less dense than
the average density of the liquid. Note also that the decagon number varies
with temperature even within the stability region of the quasicrystal. The
reason for this behavior is not the presence of structural defects
(e.g.\ vacancies), but a change in the tile occurrence ratio.

The static structure factor (Fig.~\ref{fig:dec_phase_img}, right)
shows the ten-fold symmetry of the quasicrystal. Strong diffuse
scattering surrounds sharp peaks. According to the theorem by Mermin
and Wagner~\cite{MerminWagner}, Bragg peaks are not possible in two
dimensions, but only algebraic divergencies. A close analysis of the
figure reveals that the peaks (especially the inner weak peaks) are
not aligned perfectly. The reason is the presence of residual phason
strain in the system that has not relaxed completely over the finite
simulation time. When comparing independent simulation runs, we find
that the phason strain varies from simulation to simulation. In
contrast, during the course of a single longer simulation, only little
variation is observed after the initial crystallization. This suggests
that the presence of phason strain is a finite-size effect. We expect
that the usage of larger simulation boxes would in average lead to
smaller phason strain.

\begin{figure}
\begin{center}
\includegraphics[angle=-90,width=\columnwidth]{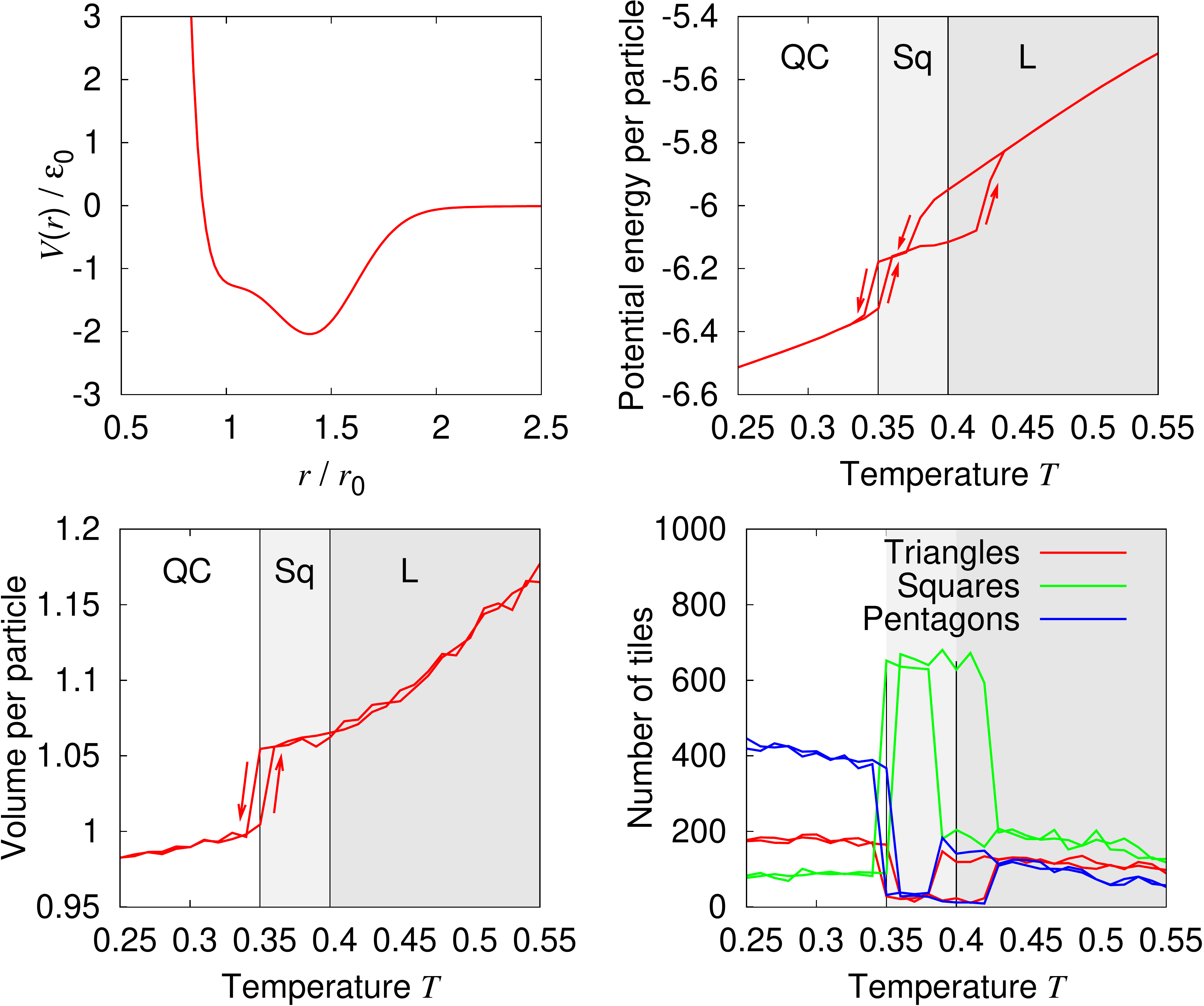}
\end{center}
\vspace*{-5mm}
\caption{ (Top left)~The Lennard-Jones-Gauss potential for $r_G=1.42$,
  $\varepsilon=1.8$, and $\sigma^2=0.042$ is a single well with shoulder.
  (Top right)~The system exhibits two first order phase transitions with
  (Bottom left)~positive thermal expansion and no thermal expansion,
  respectively. (Bottom right)~While the quasicrystal~(QC) and the liquid~(L)
  have triangular, square-like, and pentagonal local configurations, no
  triangles or pentagons are present in the square phase~(Sq).}
\label{fig:pote_dodec2}
\end{figure}

A dodecagonal quasicrystal is found for the parameters
\begin{align} \label{parameter:dodec2}
r_G=1.42\text{, } \varepsilon = 1.8\text{, and } \sigma^2=0.042.
\end{align}
The potential energy and the specific volume as a function of temperature
(Fig.~\ref{fig:pote_dodec2}) indicate two phase transitions, the first from
the decagonal quasicrystal to a crystalline square phase at
$T_{C}=0.35\pm0.01$ and the second from the square phase to a liquid at
$T_{M2}=0.40\pm0.02$. Hysteresis appears in the potential energy for both phase
transitions. Contrary to the decagonal quasicrystal, the transition from the
dodecagonal phase to the square crystal involves (conventional) positive
thermal expansion. Neither expansion nor compression is observed for the
square-liquid transition.

\begin{figure}
\begin{center}
\includegraphics[width=0.48\columnwidth]{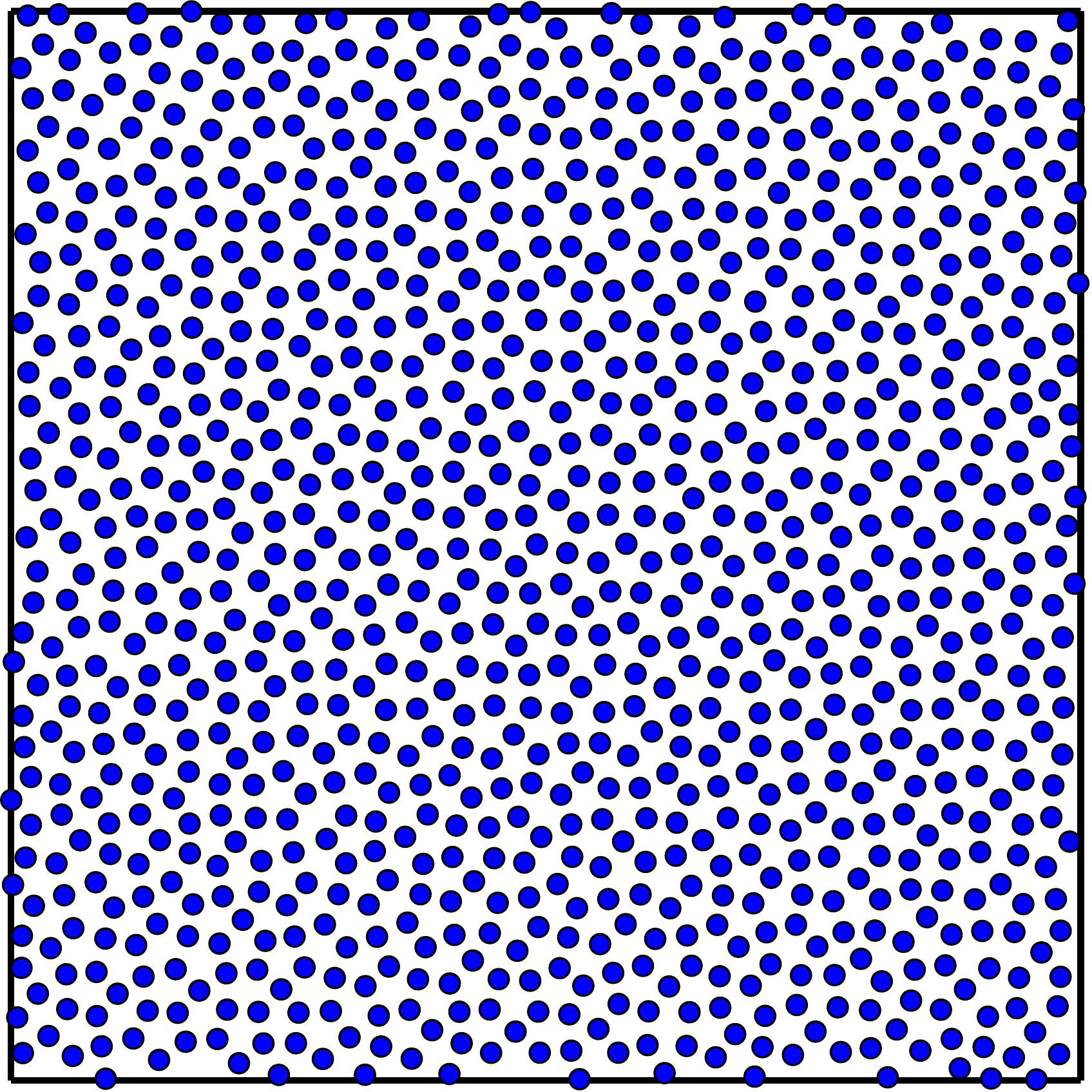}\quad
\includegraphics[width=0.48\columnwidth]{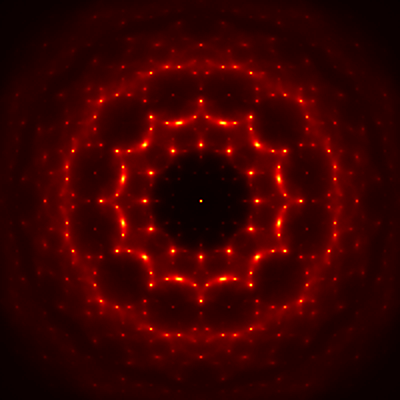}\\[0.2cm]
\includegraphics[width=0.48\columnwidth]{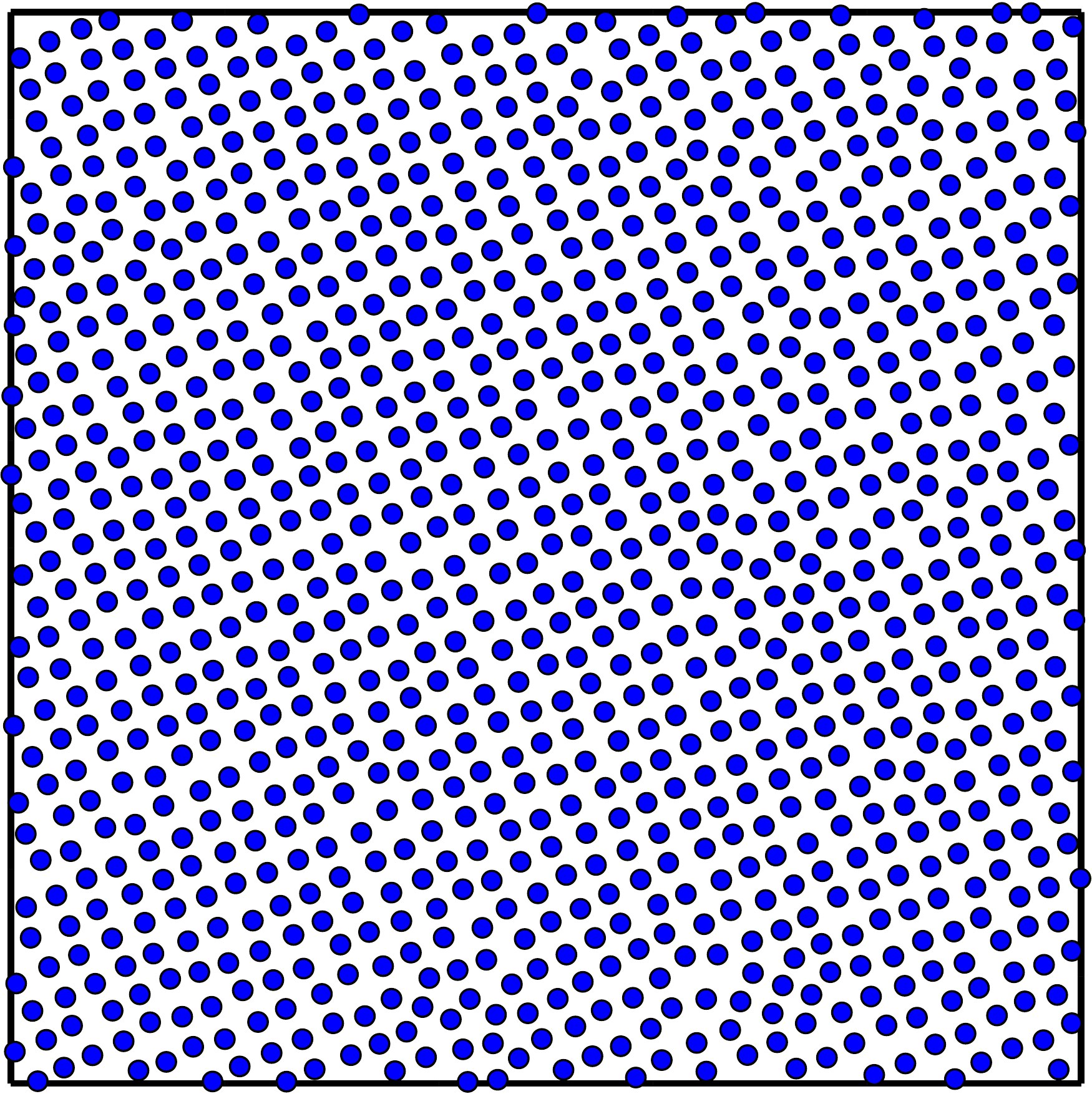}\quad
\includegraphics[width=0.48\columnwidth]{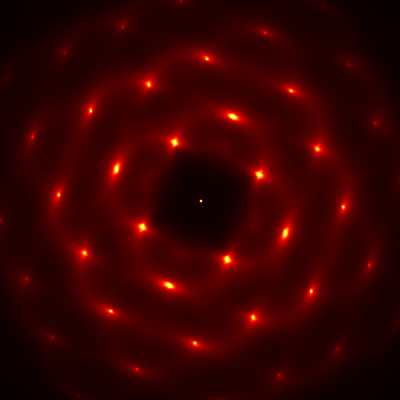}
\end{center}
\vspace*{-5mm}
\caption{Dodecagonal phase at $T=0.76T_{M2}=0.3$~(top) and square phase at
  $T=0.94T_M=0.37$~(bottom). (Left)~Particle configuration in real
  space. (Right)~Intensity map of the static structure factor.}
\label{fig:dodec2_phase_img}
\end{figure}

Only three tile types appear in the dodecagonal quasicrystal
(Fig.~\ref{fig:dodec2_phase_img}, top): triangles, squares, and
pentagons. Most pentagons are surrounded by a ring of twelve particles, but
the five-fold symmetry of the pentagons breaks the twelve-fold symmetry. The
square crystal (Fig.~\ref{fig:dodec2_phase_img}, bottom) has no triangles or
pentagons. As will be shown below, particles are highly mobile in the square
phase. This explains the appearance of a small number of defects (pentagons)
found in the figure, which are present in thermodynamic equilibrium and help
to stabilize the square phase (see the discussion in Sec.~\ref{propensity}).

The static structure factors for the dodecagonal quasicrystal and the
square phase show twelve-fold and four-fold symmetry, respectively
(Fig.~\ref{fig:dodec2_phase_img}, right). Rings of diffuse scattering
are observed in the dodecagonal phase. As in the decagonal phase, the
weak peaks are not well aligned, which is a hint for the presence of
phason strain. The square crystal also shows pronounced diffuse
scattering in the background, but in contrast to the quasicrystal the
scattering appears in streaks instead of rings. The presence of such
strong diffuse scattering in the square crystal is unusual and
indicates a large local mobility of the particles in the square phase.

\section{Statistics of particle flips}\label{statistics}

In this section, we show that the elementary processes for displacive particle
dynamics are single-particle flips in case of the decagonal quasicrystal and
multi-particle flips in case of the dodecagonal quasicrystal. A sequence of
flips is needed for collective motion as necessary for diffusion.

\subsection{Individual flips}

The average single-particle dynamics in a solid is measured by the van
Hove autocorrelation function $G_{a}(\mathbf{r},t)$. Particles do not
move far over short times, and the dominant contribution to $G_{a}$
comes from thermal motion around the local potential energy minima
closest to the particle position. The oscillatory motion around the
local equilibrium positions results in a peak at $\mathbf{r}=0$. In
case of a harmonic well or at low temperatures, the peak has Gaussian
shape.

\begin{figure}
\begin{center}
\includegraphics[width=\columnwidth]{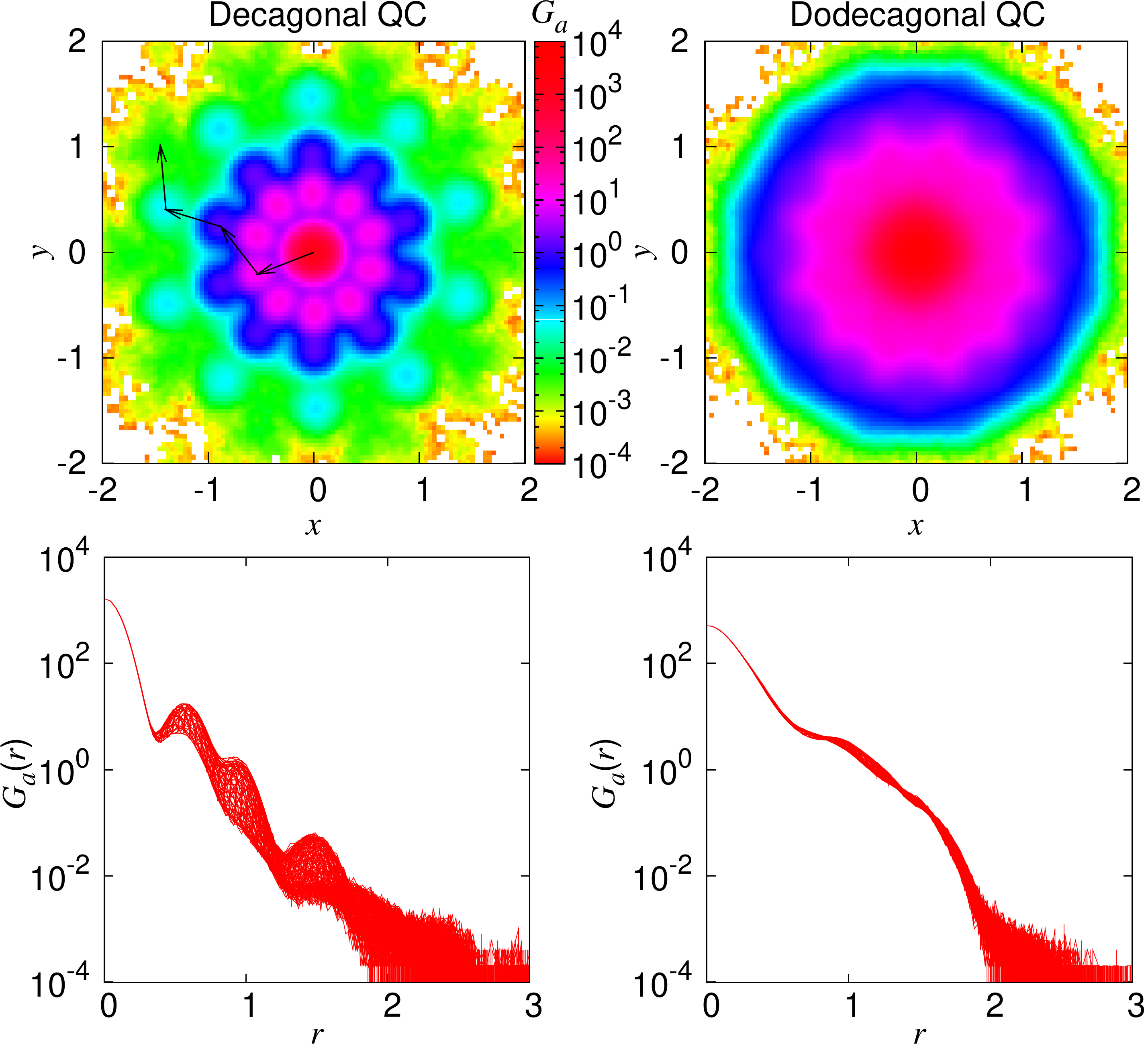}
\end{center}
\vspace*{-5mm}
\caption{The Van Hove autocorrelation function $G_{a}(\mathbf{r},t)$,
  $\mathbf{r}=(x,y)$, $t=100$ for the decagonal quasicrystal at
  $T=0.5$~(left) and the dodecagonal quasicrystal at $T=0.3$~(right)
  is strongly broadened, highly anisotropic~(top), and decays exponentially
  with $r$~(bottom).}
\label{fig:Gself_dec}
\end{figure}

The van Hove autocorrelation function $G_{a}(\mathbf{r},t)$,
$\mathbf{r}=(x,y)$ is shown in Fig.~\ref{fig:Gself_dec}. We fix the
time interval to $t=100$, which is two to three orders of magnitude
larger than the typical time for oscillation in a local minimum
($\Delta t=0.1$ to 1.0). $G_{a}$ is non-isotropic for both
quasicrystals and deviates from the expected behavior for a
crystalline solid. In the case of the decagonal phase, rings of local
maxima with outwards decreasing heights surround a central peak. The
central peak of the decagonal quasicrystal decays exponentially and is
significantly broader than what would be expected from pure
oscillatory motion. This indicates that particles are highly mobile
and phonons are not the only mechanism of particle dynamics.

\begin{figure}
\begin{center}
\includegraphics[width=\columnwidth]{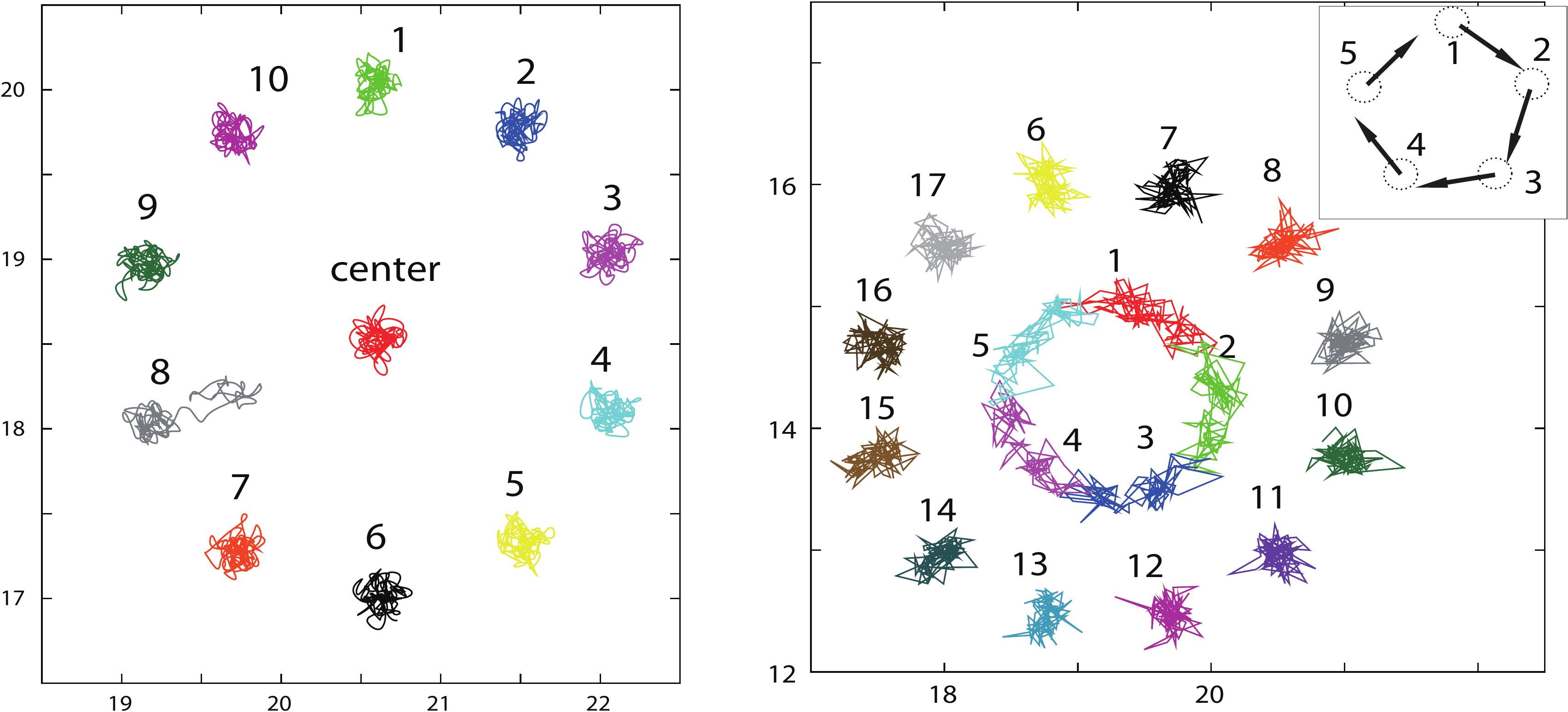}
\end{center}
\vspace*{-5mm}
\caption{Trajectories of particles in high-symmetry clusters ($\Delta t =
  10$). (Left)~Eleven particles form a decagon cluster. Particle `8' makes a
  single-particle flip. (Right)~17 particle form a dodecagon
  cluster. Multi-particle flip involve the five inner particles.}
\label{fig:trajec_small11}
\end{figure}

To investigate the unusually high particle mobility in the quasicrystals, we
study trajectories of particles belonging to characteristic high-symmetry
clusters. A decagon cluster (Fig.~\ref{fig:trajec_small11}, left) consists of
a single central particle surrounded by a ring of ten particles. In the
figure, one of the particles (`8') moves away from its position at a ring
vertex to a position in the interior of the ring where it remains for the rest
of the trajectory. The time for the switch of position is short, circa $\Delta
t=0.3$, which is in the order of a single phonon oscillation. We can say that
the particle `jumped' to its new position. The motion is an example of a
\emph{single-particle flip}.

The dodecagon cluster (Fig.~\ref{fig:trajec_small11}, right) consists
of five central particles surrounded by a ring of twelve particles. As
shown by the trajectories of the particles, the elementary dynamical
process is the correlated rotation of the central particles called a
\emph{multi-particle flip}. The rotary motion is not as well-defined
as the single-particle flip. Since the five-fold center breaks the
symmetry of the twelve-fold ring, the multi-particle flip has a lower
energy barrier than the single-particle flip in the decagonal phase.

We can determine the flip distances and flip directions of the
decagonal phase from the peaks positions in
Fig.~\ref{fig:Gself_dec}. There is only a single flip distance:
$\Delta r\approx 0.6$. In an ideal tiling of edge length 1, the flip
distance would be $\tau^{-1}\approx 0.62$ with the golden mean
$\tau=(\sqrt{5}+1)/2$. The sequence of arrows in
Fig.~\ref{fig:Gself_dec}(left) indicate consecutive flips: An example
of a single flip is the jump to the end of the first arrow, an example
of two second flip is the jump to the end of the second arrow, etc. Up
to four consecutive flips are observed during the observation time of
$\Delta t = 10$. As expected, the probability for consecutive flips
decays rapidly with the number of flips.

\begin{figure}
\begin{center}
\includegraphics[width=\columnwidth]{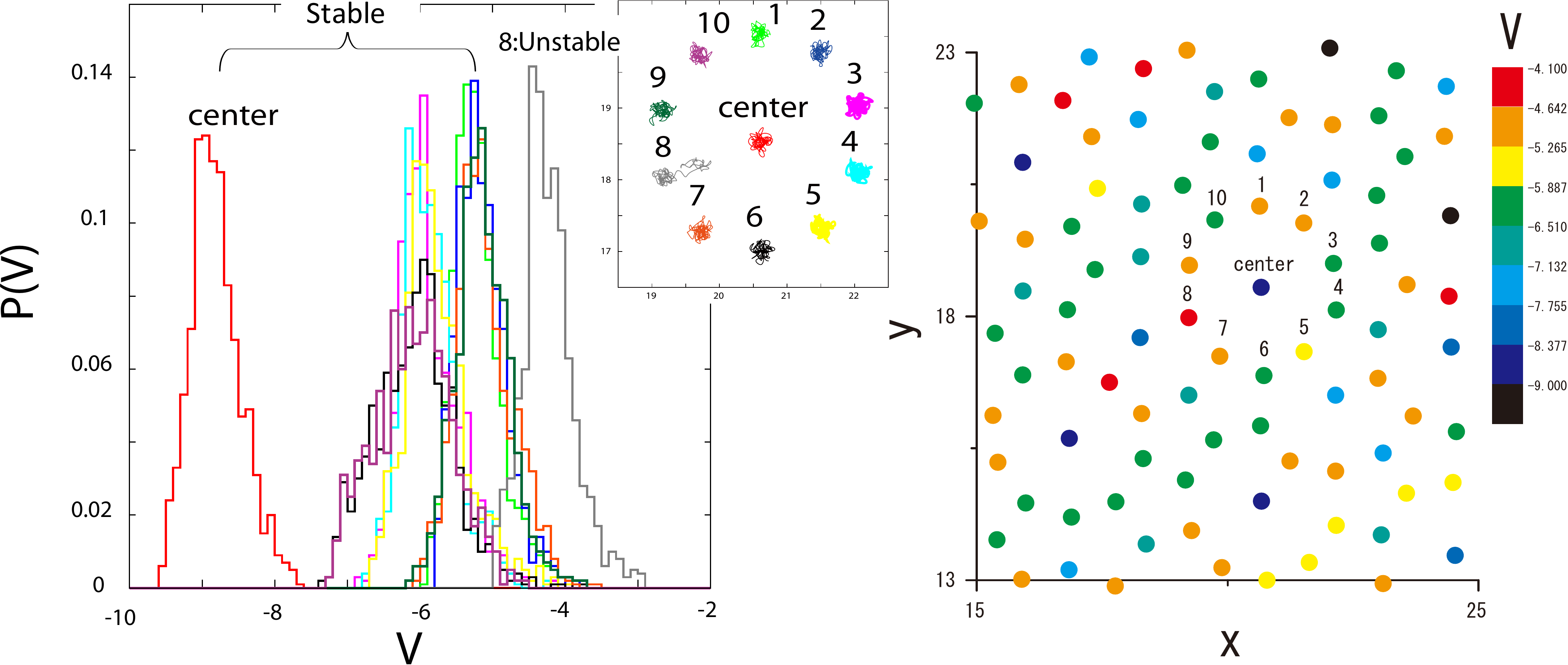}
\end{center}
\vspace*{-5mm}
\caption{(Left) Potential energy histograms for the particles in the
  decagon cluster of Fig.~\ref{fig:trajec_small11}(left) averaged over
  the time $\Delta t=10$.  (Right) Average potential energies of
  individual particles.}
\label{fig:histogram_dec}
\end{figure}

What determines which of the particles is going to flip next? Besides
geometric restrictions, which can be understood from the underlying
tiling~\cite{EngelStability}, the height of the energy barrier plays an
important role. Equilibrium positions with higher potential energies will in
general be less favorable. Potential energy histograms for each of the eleven
particles of the decagonal cluster are shown in Fig.~\ref{fig:histogram_dec}.
The central particle has by far the lowest energy with
$E_{\text{center}}\approx-9$. The particles on the decagon rings have energies
$E_{j}\approx-6$ for $j=3,4,5,6,10$, $E_{j}\approx-5$ for $j=1,2,7,9$, and
$E_{j}\approx-4$ for $j=8$. The reason for the appearance of three energies is
the different local configuration of the particle on the outside of the
ring. The particle with highest potential energy is the one that is observed
to flip in Fig.~\ref{fig:trajec_small11}(left).

\subsection{Collective flips}

\begin{figure}
\begin{center}
\includegraphics[width=\columnwidth]{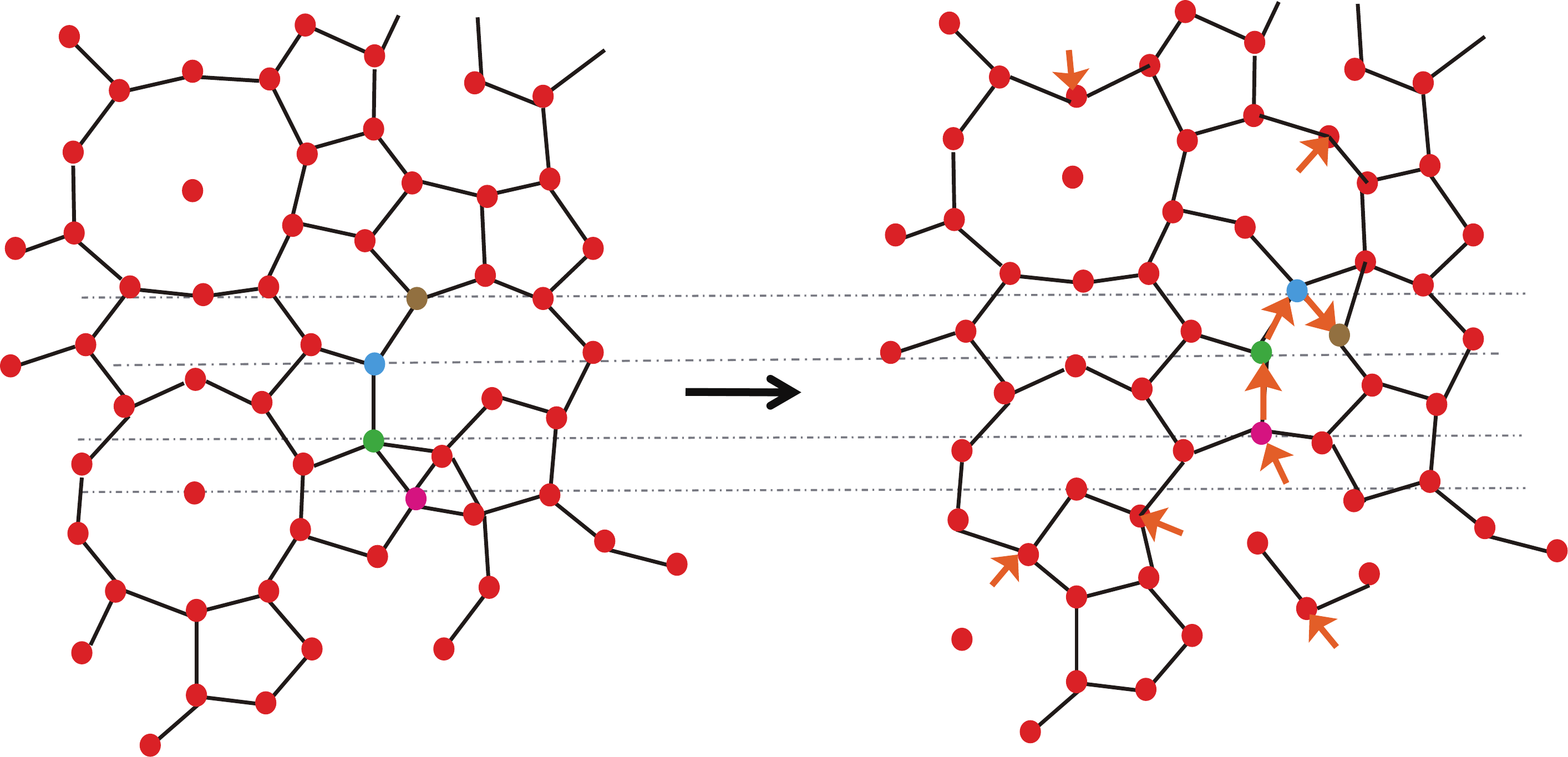}
\end{center}
\vspace*{-5mm}
\caption{Example of phason flips in the decagonal quasicrystal. The
  configuration on the left side transforms into the configuration on the
  right side. Orange arrows are the trajectories of the flipping atoms.  The
  pink, green, blue, and brown particles perform a string-like sequence of
  flips.}
\label{fig:collective_flips}
\end{figure}

Individual flips transform one local configuration into a similar,
energetically nearly degenerate one. Only if consecutive flips can generate
structural changes that are flexible enough, then they can induce a collective
reorganization of the quasicrystal and lead to particle motion over long
distances and eventually diffusion. An example for consecutive flips in the
decagonal phase is shown in Fig.~\ref{fig:collective_flips}. Two particles
(red points) are connected by black lines if they are neighbors, i.e.\ if
their distance corresponds to the first peak in the radial distribution
function. The left panel is the initial configuration, and the right figure
the final configuration after the flips have occurred. The displacement of the
flipping atoms is indicated by arrows. Dashed lines serve as a guide for the
eyes to identify the particle motion. Five individual flips (single arrow) and
a string-like chain of four consecutive flips (four arrows in the center) can
be seen.
 
\begin{figure}
\begin{center}
\includegraphics[width=\columnwidth]{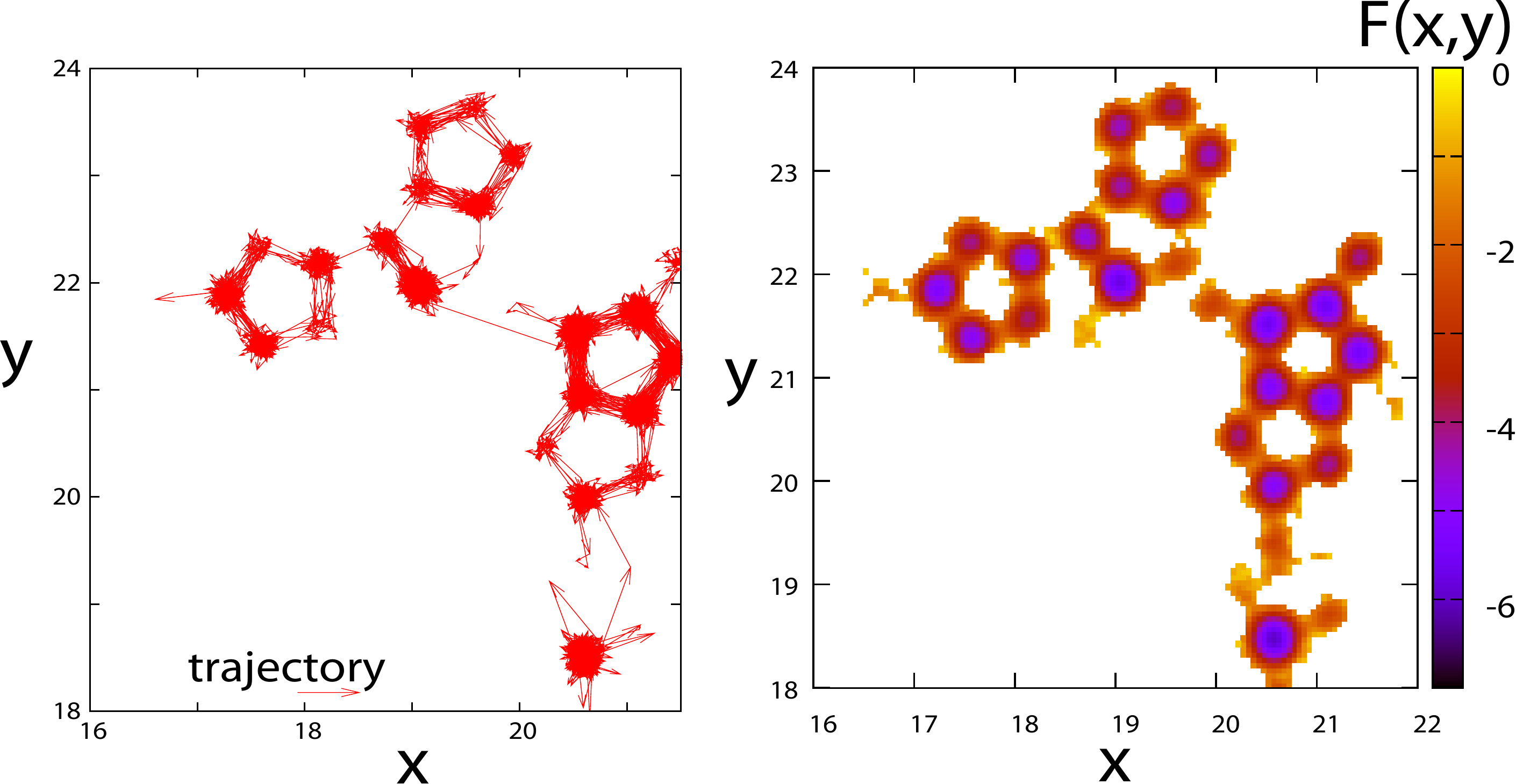}
\end{center}
\vspace*{-5mm}
\caption{(Left)~Trajectory of a single particle in the decagonal
  quasicrystal. The position is recorded every $\Delta t = 10$ over a total
  time of $t=10^5$. The particle exhibits diffusive motion along the vertices
  of pentagons. (Right)~The effective free energy of the particle as
  calculated from the probability density of the particle position.}
\label{fig:trajec_long}
\end{figure}

The long-time dynamics of a single particle is analyzed in
Fig.~\ref{fig:trajec_long}. The particle under investigation is the `center'
particle in Fig.~\ref{fig:trajec_small11}(left). Its trajectory is recorded over a
total time of $t=10^5$. At the beginning, the particle remains stable for some
time, because it starts at a low potential energy position
(Fig.~\ref{fig:histogram_dec}). In the time range $t < 10^4$, few
back-and-forth jumps to neighboring positions are observed. Only later, the
particle starts moving further away. Equilibrium positions frequently form
pentagons on the intermediate time scale. The time for a switch from one
pentagon to another is on the order of $\Delta t=10^4$ at the temperature
under investigation ($T=0.5$).

We can define an effective free energy $F(x,y)=-k_B T\ln Z(x,y)$ for a
particle positioned at $(x,y)$. The restricted partition function $Z(x,y)$ is
obtained by averaging over a constraint phase space, where the selected
particle is tagged at $(x,y)$ and the others are allowed to perform only
phonon motion but no additional flips. $F(x,y)$ can then be interpreted as the
free energy landscape for this particle~\cite{FreeEneLandscape}. In simulation,
the effective free energy can be determined in a first approximation from the
probability density $P(x,y)$ of the particle position:
\begin{align} \label{eq:free_ene}
F(x,y)=-k_BT \log P(x,y).
\end{align}
Fig.~\ref{fig:trajec_long}(right) shows the intensity map of $F(x,y)$
for the particle on the left side. The particle favors staying at
equilibriums positions arranged on vertices of pentagons.  The
particle dynamics can be understood as a flip motion among the basins
of the free energy landscape.

\subsection{Diffusion}

The diffusivity $D(T)$ of the quasicrystals can be determined from the
average slope of the mean square displacement $\langle
r^2(t)\rangle$. In the case of the decagonal quasicrystal, we also
measure the flip frequency as a function of temperature by counting
the number of flips during the simulation.

\begin{figure}
\begin{center}
\includegraphics[width=\columnwidth]{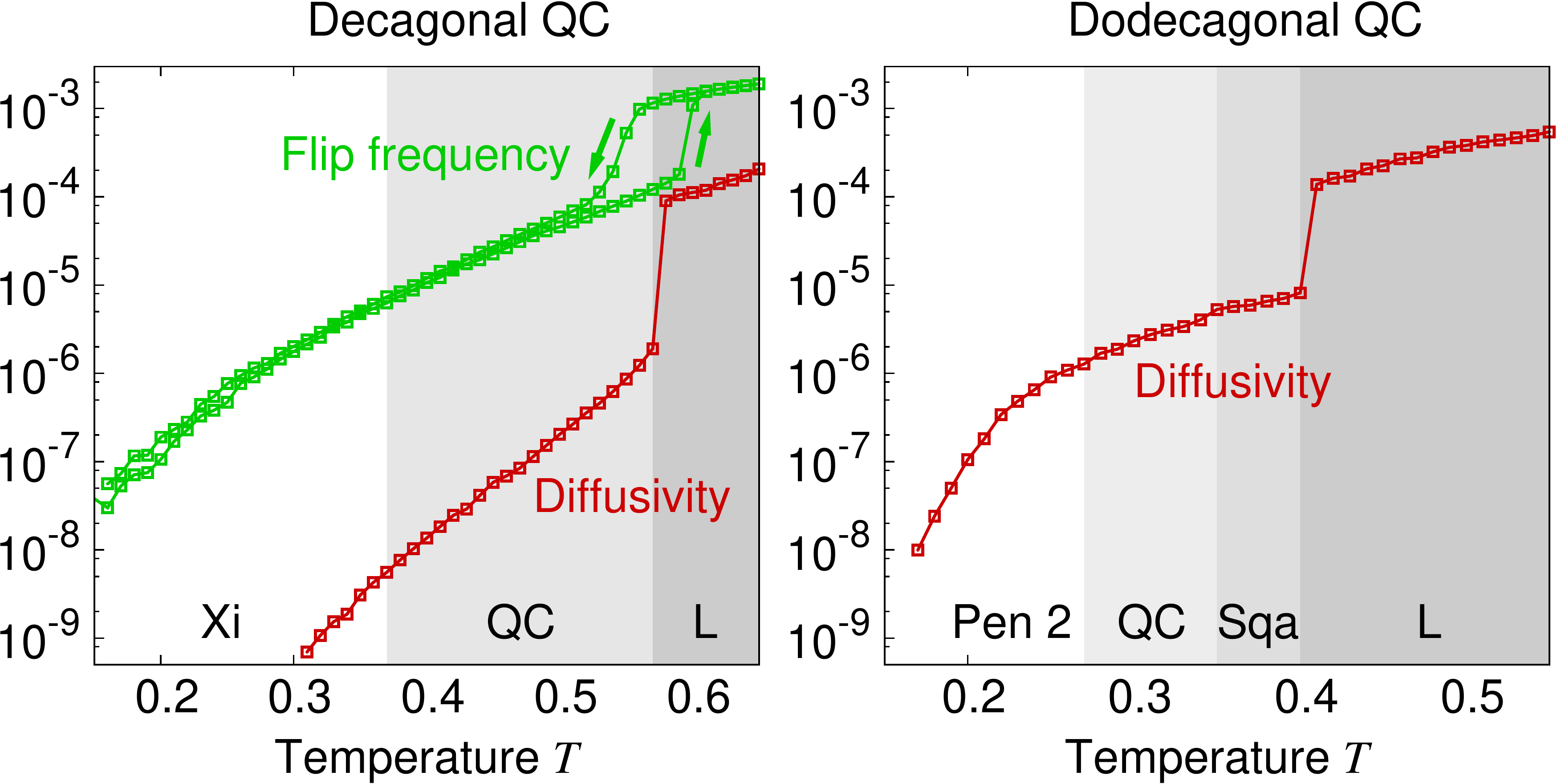}\\
\vspace*{2mm}
\includegraphics[width=\columnwidth]{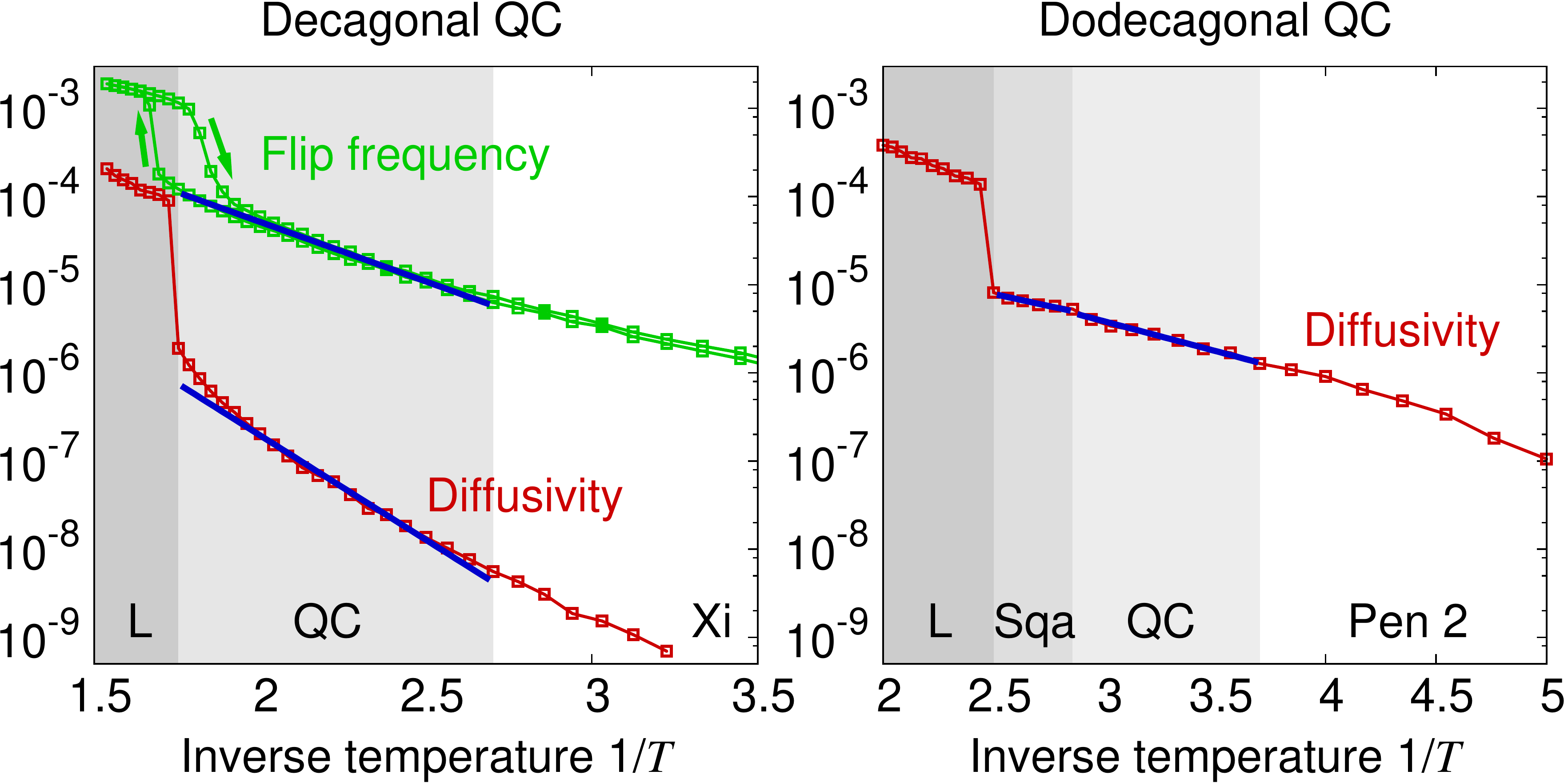}
\end{center}
\vspace*{-5mm}
\caption{The diffusivity and the flip frequency of the decagonal
  quasicrystal~(Left) and the diffusivity of the dodecagonal
  quasicrystal~(Right) as a function of
  temperature~(Top). (Bottom)~The temperature dependence can be fitted
  with an Arrhenius law as indicated by blue lines. At low
  temperatures and in the absence of finite-size effects, the
  quasicrystals transform into the periodic approximants Xi and Pen
  2.}
\label{fig:diffusion}
\end{figure}

We find that the diffusivity and the flip frequency decreases rapidly
over many orders of magnitude during lowering of temperature
(Fig.~\ref{fig:diffusion}). In order to investigate whether individual
flips and diffusion are activated processes, we fit the curves by an
Arrhenius law
\begin{align}
D(T)=D_0\exp(-\Delta E/k_BT).
\end{align}
The fit, indicated by blue lines in Fig.~\ref{fig:diffusion}, works
well confirming that flips and diffusion are indeed dominated by
energy barriers. The slopes for the curves correspond to activation
energies $\Delta E$. Activation energies for diffusion in the
decagonal quasicrystal (`10'), the dodecagonal quasicrystal (`12'),
the square crystal (`4') and for flips in the decagonal quasicrystal
(`f') are:
\begin{align}
\Delta E_{10}=5.46(14),\\
\Delta E_f=3.11(10),\\
\Delta E_{4}=1.26(18),\\
\Delta E_{12}=1.54(7).
\end{align}
The small deviations from the Arrhenius law observed in
Fig.~\ref{fig:diffusion} for low and high temperatures can be a result
of the change of barrier heights with temperature, which we expect to
happen due to particle interactions during a flip. For example, if a
particle wants to squeeze through two other particles to achieve a
flip, then it is possible that the latter particles are pushed to the
sides to make it easier for the former particle to pass
through. Cooperative effects like this one will change with
temperature, which alters the effective barrier heights. Furthermore,
the change of the tile occurrence ratio with temperature can modify the
flip type and therefore energy barrier height distribution. Finally,
defect tiles and vacancies can be present at higher temperature and
constitute a competing mechanism for diffusion, which can lead to a
deviation from the ideal Arrhenius law. As can be seen in
Fig.~\ref{fig:diffusion} (bottom), the effective energy barrier
heights increase with increasing temperature for the decagonal system,
while they decrease with increasing temperature for the dodecagonal
system.

The activation energy $E_f$ calculated for the flip frequency in the
decagonal phase is lower than the activation energy $E_{10}$
calculated for the diffusivity. Since diffusion takes place by
successive flips, there has to be a distribution of energy barrier
heights for individual flips. In general, lower barriers lead to
faster movement. The highest barrier that has to be overcome for
long-range particle motion to occur constitutes the bottleneck for
diffusion.

Variations of barrier heights can be seen indirectly in
Fig.~\ref{fig:trajec_long}. It is observed that the particle motion
along the vertices of pentagons (`intra-pentagon flip') is fast and
therefore corresponds to relatively low barriers. On the other hand,
the transition from one pentagon to another one (`extra-pentagon
flip') is found to be much slower and corresponds to relatively high
barriers. The ratio of times for the two types of motions is
approximated by $t_\text{ep}/t_\text{ip}=\exp((\Delta E_{10}-\Delta
E_\text{f})/k_BT)$. For $T=0.5$, $t_\text{ep}/t_\text{ip}\approx 100$.
This agrees well with what is typically observed in simulation. In
other words, for every 100 intra-pentagon flips there is about one
extra-pentagon flip, lowering the diffusivity compared to what would
be expected from the flip frequency alone. Extra-pentagon flips are
the bottleneck for long-range diffusion in the decagonal quasicrystal.

The energy barrier for diffusion in the dodecagonal quasicrystal is
lower than the energy barrier for diffusion in the decagonal
quasicrystal.  This means that the dodecagonal quasicrystal
reorganized more easily. The single-well nature of the LJG potential
for the dodecagonal system helps to facilitate flips and explains the
low energy barriers. In contrast, the intermediate bump in the LJG
potential for the decagonal system leads to higher energy barriers.

As we will see below, the reason for the fast diffusion in the square
crystal is of different nature; it is connected to the formation and
propagation of local defects. Such defects are easily possible due to
the softness of the interaction potential.

Note that both quasicrystals are stabilized entropically only above a
certain critical temperature; at lower temperatures they transform
into periodic approximants, i.e.\ periodic crystals with a similar
local structure as the quasicrystals. The decagonal quasicrystal is
unstable relative to the approximant Xi for $T\leq0.37$~\cite{LJG};
and the dodecagonal quasicrystal is unstable relative to the
approximant Pen 2 for $T\leq0.25$~\cite{Engel_Dpaper}. In our
simulations, periodic boundary conditions suppress the transformations
from quasicrystals to approximants. Nevertheless, we do observe that
the dynamics slows down in the regions where the approximants are
stable (most prominently for the dodecagonal quasicrystal) as visible
by the deviation from the Arrhenius law at low temperatures in in
Fig.~\ref{fig:diffusion}. Here, the quasicrystals are unstable.

\section{Dynamic propensity}
\label{propensity}

\begin{figure}
\begin{center}
\includegraphics[width=0.48\columnwidth]{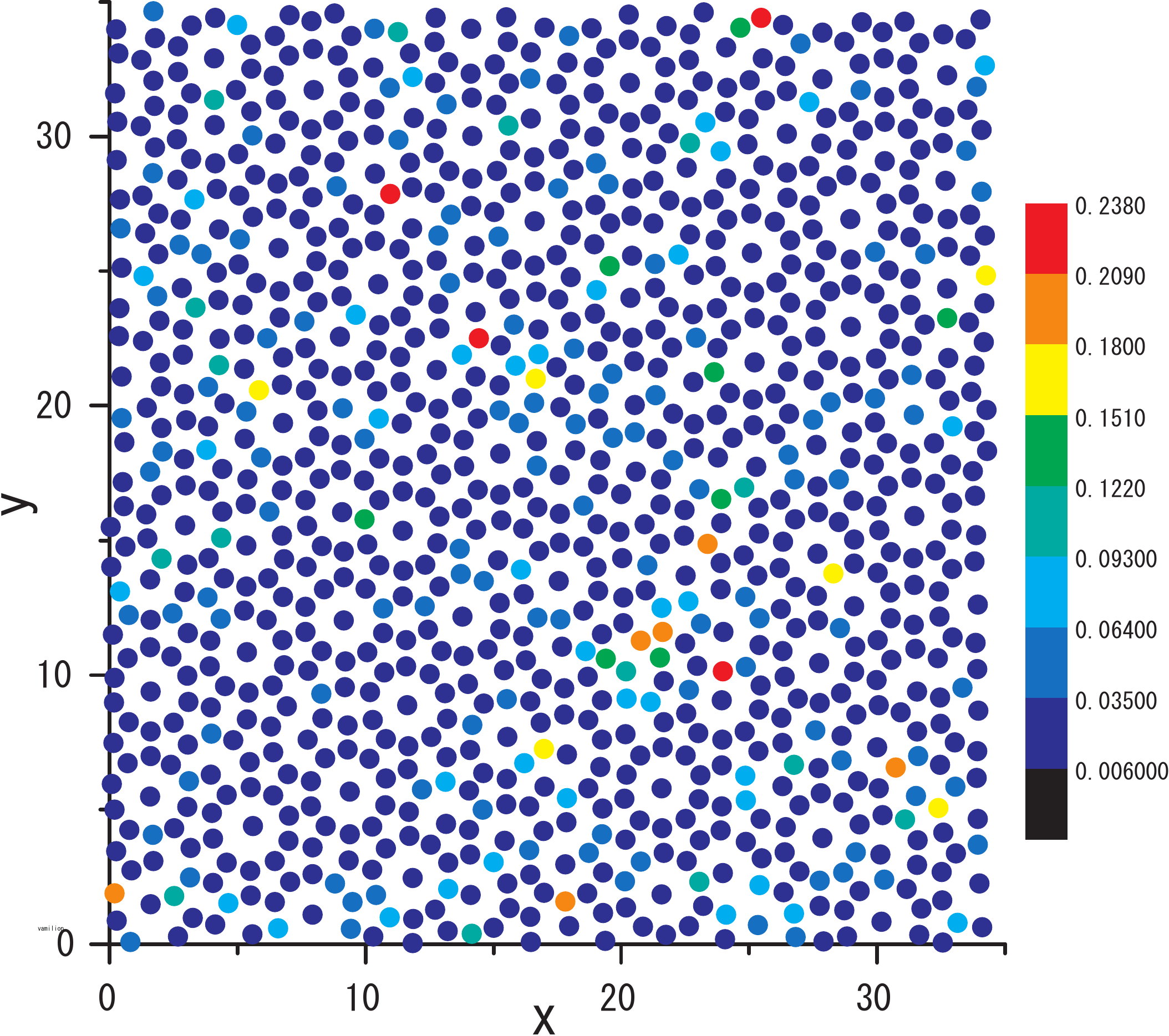}
\end{center}
\vspace*{-5mm}
\caption{Spatial distribution of the propensity $\phi_j(t)$, $t=10$ for the
  decagonal phase at $T=0.5$. The centers of the circles are atomic positions,
  and the color represents propensity.  Red means high propensity, and dark
  blue low propensity.}
\label{fig:dec_propensity}
\end{figure}

To investigate the distribution of mobile particles, the temperature
dependence of the dynamic propensity $\phi_j(t)$ is
studied~\cite{propensity1,propensity2}. Since we are interested in short time
motion, we fix $t=10$ in the following. The iso-configurational average (see
Eq.(\ref{eq:propensity})) is taken over 1000 different initial
velocities. Fig.~\ref{fig:dec_propensity} illustrates the spatial distribution
of $\phi_j(t)$ at $T=0.5$ in the decagonal phase. We observe that most of the
particles have low mobility. Particles with high mobility appear predominantly
at the edges of decagon clusters. From the propensity of the green, yellow,
and red particles ($\phi_j>0.09$) we estimate that these particles move in
average over a distance of $\Delta r>0.3$, which means they perform flips with
a probability of more than 50\%.

\begin{figure}
\begin{center}
\includegraphics[width=\columnwidth]{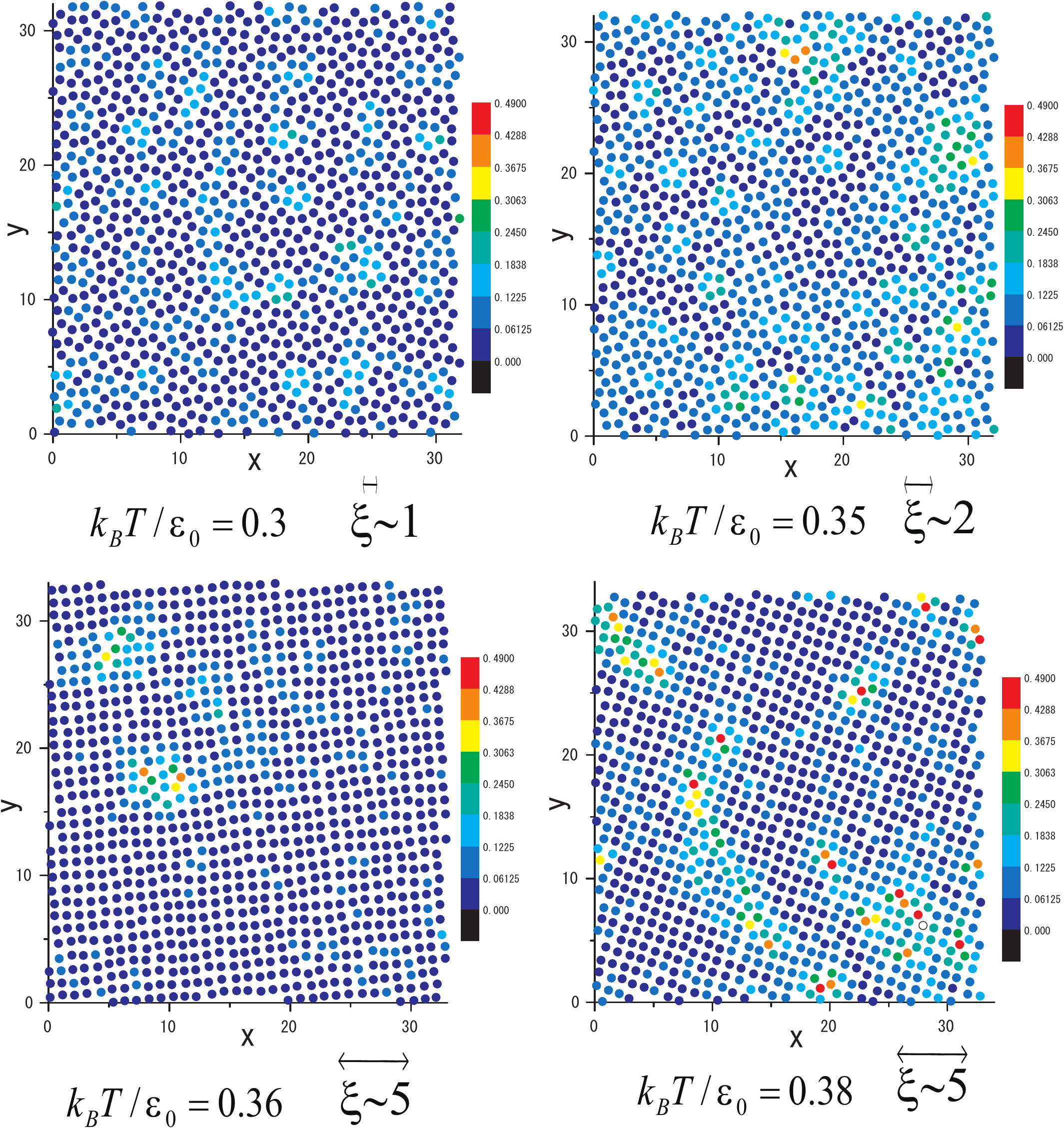}
\end{center}
\vspace*{-5mm}
\caption{Spatial distribution of the propensity $\phi_j(t)$, $t=10$ for the
  dodecagonal phase at $T=0.5$. The centers of the circles are atomic
  positions, and the color represents propensity. The decagonal quasicrystal
  at $T=0.3$~(top left) and $T=0.35$~(top right), and the square crystal at
  $T=0.36$~(bottom left) and $T=0.38$~(bottom right) is shown. The dynamical
  correlation length $\xi$ is given below the subfigures.}
\label{fig:dodec2_propensity}
\end{figure}

Next, we consider the propensity in the case of the dodecagonal system, in
particular close to the quasicrystal-square crystal transition at $T_{C}=0.35$
and close to the melting transition at
$T_{M2}=0.4$. Fig.~\ref{fig:dodec2_propensity} shows the spatial distribution
of propensity at the temperatures $T=$0.3, 0.35, 0.36, and 0.38. In the
dodecagonal phase ($T=0.3$), the five central particles of dodecagonal
clusters often exhibit high mobility.  When the temperature is increased
($T=0.35$), mobility increases and larger areas are observed to have high
propensity. In the square phase ($T=0.36$), defect sites of high mobility are
pentagonal rings.  As the temperature is increased ($T=0.38$), the defect
domains grow.

\begin{figure}
\begin{center}
\includegraphics[width=0.5\columnwidth]{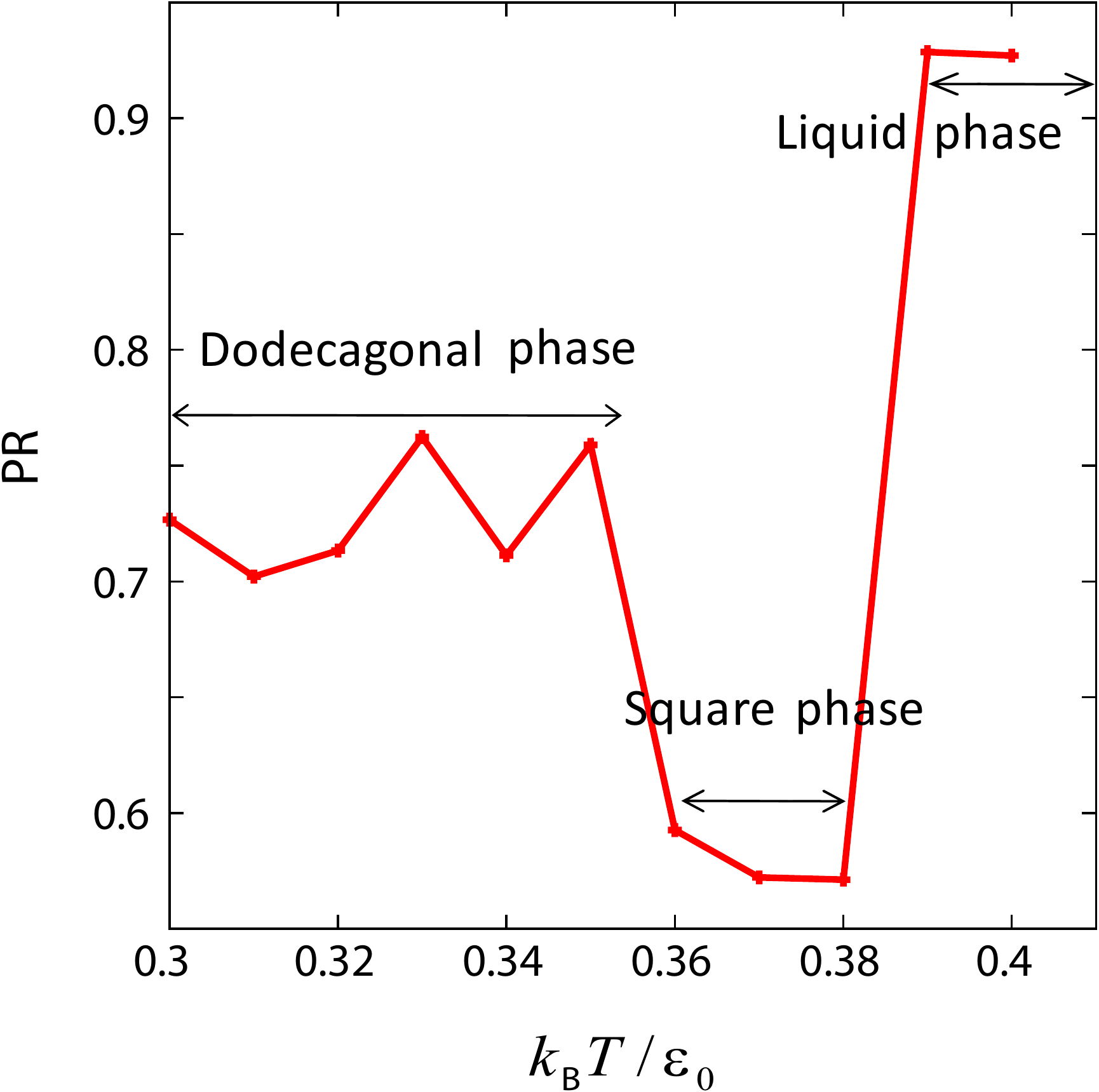}
\end{center}
\vspace*{-5mm}
\caption{Temperature dependence of the participation ratio in the
  dodecagonal system. The low value in the square phase indicates particle
  dynamics dominated by structural defects.}
\label{fig:PR}
\end{figure}

To analyze the size of the mobile regions, we calculate the participation
ratio (PR) of the propensity for the dodecagonal system. The participation
ratio is defined by
\begin{align}
\text{PR}&=\dfrac{1}{N}\dfrac{\left ( \sum_{j=1}^N \phi_j \right )^2}
{\sum_{j=1}^N \phi_j^2}.
\end{align}
Note that the participation ratio satisfies $\text{PR}\approx 1$ for extended
states and $\text{PR}\ll 1$ for localized
states~\cite{Primer}. Fig.~\ref{fig:PR} shows the temperature dependence
for $0.3\leq T\leq 0.4$. We can identify three regimes: For $0.39\leq T$
(liquid), the participation ratio is close to unity, which indicates that
mobile particles are distributed homogeneously in the sample. The
participation ratio for $0.35<T<0.39$ (square crystal) is lower than that for
$T\leq0.35$. This is a characteristics of localized defects. The higher
participation ratio for $T<0.35$ (quasicrystal) indicates that mobile
particles are again distributed regularly in real space.

\begin{figure}
\begin{center}
\includegraphics[width=\columnwidth]{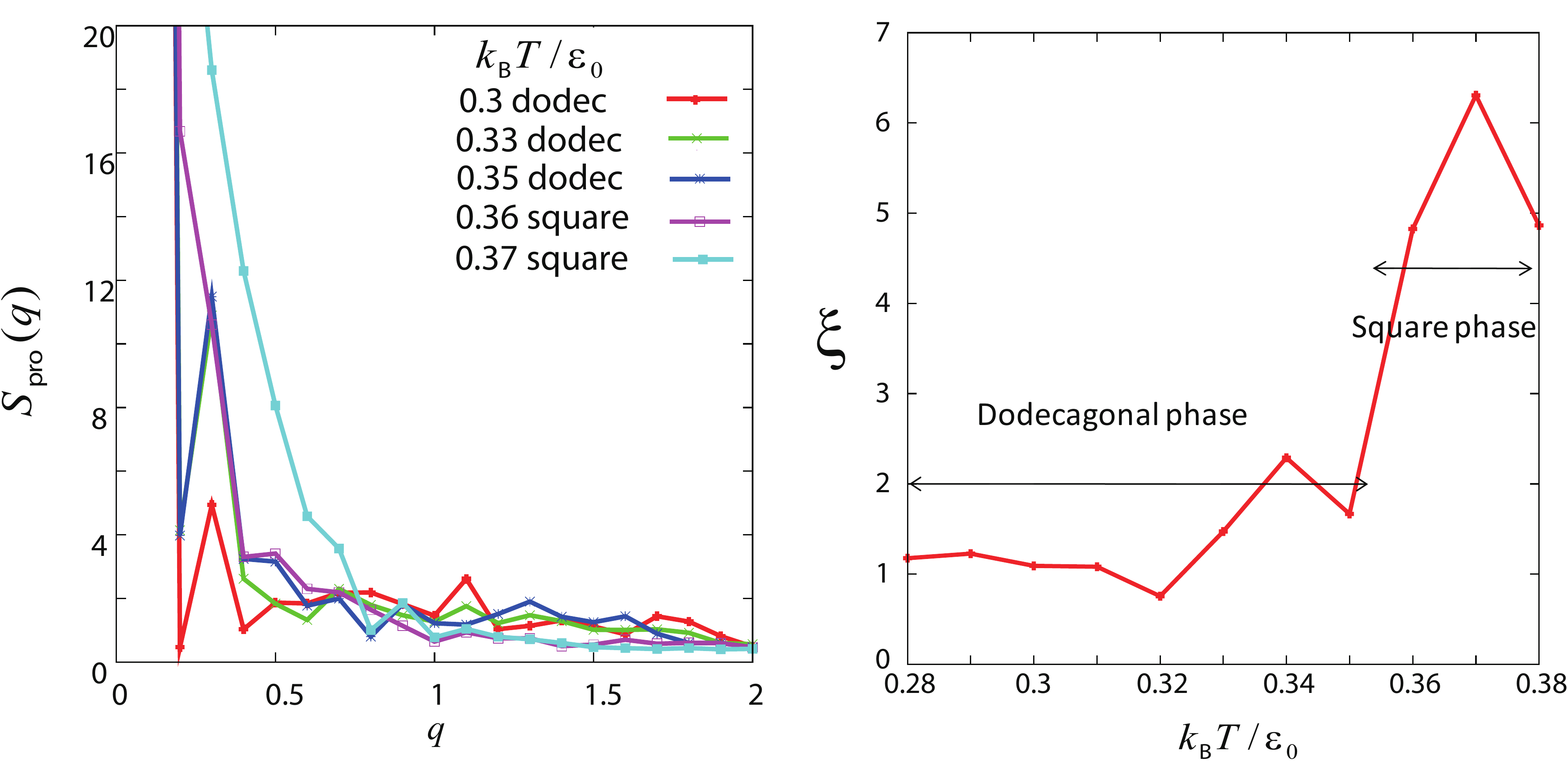}
\end{center}
\vspace*{-5mm}
\caption{(Left)~Fourier transformed propensity distribution
  $S_{\text{pro}}(q)$ at various temperatures. (Right)~Temperature dependence
  of the dynamical correlation length $\xi(T)$ in the dodecagonal
  quasicrystalline phase and the square phase.}
\label{fig:pro_Sq}
\end{figure}

To measure the correlation of mobile particles, we study the Fourier
transformed propensity distribution
\begin{align}\label{eq:propensity_col}
S_{\text{pro}}(\mathbf{q})=\frac{1}{N'} \left| \sum_{\phi_j > \bar \phi}
\phi_j \exp(\dot \imath \mathbf{q}\cdot\mathbf{x}_j)
\right|^2,
\end{align}
where the sum is over particles with propensity higher than the average value
$\bar\phi = \sum_j\phi_j/N$ and $N'$ is the number of such particles.
Fig.~\ref{fig:pro_Sq}(left) shows the radial average $S_{\text{pro}}(q)$ at
various temperatures. In the range of small wave vectors, $S_{\text{pro}}(q)$
can be fitted by a Cauchy distribution:
\begin{align}\label{eq:OZtype}
S_{\text{pro}}(q)\propto \dfrac{1}{1+\xi^2 q^2},
\end{align}
where $\xi$ is a measure for the size of mobile regions and termed dynamical
correlation length~\cite{Heterogeneities}. In the dodecagonal phase, the
dynamical correlation length is short, $\xi = 1\sim 2$, which is the size of
the pentagonal ring in the dodecagon cluster (see
Fig.~\ref{fig:dodec2_propensity}, top). In the square crystal, the dynamical
correlation length is larger, $\xi = 5\sim 6$, and corresponds to the size of
the defect domains (see Fig.~\ref{fig:dodec2_propensity}, bottom).

\section{Conclusions}\label{conclusion}

We have investigated the particle dynamics of one-component
quasicrystals in two dimensions. The isotropic LJG pair potential is
used as a simple model system. A decagonal quasicrystal and a
dodecagonal quasicrystal are thermodynamically stabilized for two sets
of parameters. The growth of the quasicrystal phases from the melt
occurs via a first order phase transition with negative thermal
expansion for the decagonal quasicrystal and positive thermal
expansion for the dodecagonal quasicrystal. The static structure
factors shows a significant amount of diffuse scattering due to high
particle mobility.

The dynamics of individual particles is characterized by local
oscillatory motion (phonon dynamics) and discrete particle jumps
(phason flips), which are activated by thermal motion. We found that
an elementary flip is a single-particle jump for the decagonal
quasicrystal and a correlated ring-like multi-particle motion for the
dodecagonal quasicrystal. Due to the high structural complexity,
particles in the quasicrystals have various local environments. Phason
flips occur preferentially for those particles with potential energies
higher than the average. Over longer times, successive jumps form a
sequence of flips and particles start to diffuse through the
system. The temperature dependence of the diffusivity is well
described with an Arrhenius law, which suggests that the diffusion
mechanism is a conventional activated process. The dynamic propensity
measures the distribution of particle mobilities in the system. For
the decagonal quasicrystal, mobile particles are isolated, while for
the dodecagonal quasicrystal, pentagonal rings constitute the
dynamically active sites.

The dodecagonal quasicrystal transforms into a periodic square crystal
at increased temperatures. This is surprising at first, because
quasicrystals are assumed to have high entropy and therefore should be
increasingly preferred at higher temperatures. However, in the case of
the square phase, the lack of flip entropy is compensated by the
possibility of pentagonal structural defects, which are present in
thermodynamic equilibrium and add to the configurational entropy. In
fact, the mobility of the square phase turns out to be higher than the
mobility in the quasicrystal as confirmed by the calculation of the
participation ratio of the dynamic propensity.

It is illustrative to compare the dynamics of quasicrystals with the
particle motion observed during the transition from a supercooled
liquid to a glass. In general, the relaxation to the glassy state does
not occur homogeneously, but heterogeneously over temporal and spatial
ranges, which is called dynamical
heterogeneity~\cite{Heterogeneities}. On a local scale, phason flips
strongly resemble the slow $\beta$ relaxation process found in glassy
materials. In fact, a long-lived glassy state can be formed by the LJG
system~\cite{Mizu_oda1} and flipping motions among local free energy
minima have been observed therein as the slow $\beta$
process~\cite{Mizu_master}.

\acknowledgments

M.E. acknowledges support from the Japanese Society for the Promotion
of Science for a stay at Kyushu University, where part of this work
was conducted. The work of T.O. was supported in part by a
Grant-in-Aid for Scientific Research (C) 19540405 from the Japanese
Ministry of Education, Culture, Sports, Science and Technology

\bibliography{2dQC}

\end{document}